\documentclass[nonacm]{acmart}

\usepackage{booktabs} 

\usepackage{subfigure}
\usepackage{graphicx}

\usepackage{url}
\usepackage{breakurl}
\usepackage{multicol}
\usepackage{fancyvrb}
\usepackage{float}
\usepackage{array}
\usepackage{multirow}
\usepackage{tabularx}
\usepackage{colortbl}
\usepackage{hhline}
\usepackage{textcomp}

\usepackage[T1]{fontenc}
\usepackage[utf8]{inputenc}
\usepackage{lmodern}

\usepackage[linesnumbered]{algorithm2e}

\usepackage{amsmath}

\setcopyright{none}

\begin{document}
\title{Exploring the Intersections of Web Science and Accessibility}

\author{Trevor Bostic, Jeff Stanley, John Higgins, Daniel Chudnov, Rachael L. Bradley Montgomery, Justin F. Brunelle}
\affiliation{
  \institution{The MITRE Corporation}
}
\email{{tbostic, jstanley, jphiggins, dlchudnov, rbradley, jbrunelle}@mitre.org}

\begin{abstract}
The web is the prominent way information is exchanged in the 21st century. However, ensuring web-based information is accessible is complicated, particularly with web applications that rely on JavaScript and other technologies to deliver and build representations; representations are often the HTML, images, or other code a server delivers for a web resource. Static representations are becoming rarer and assessing the accessibility of web-based information to ensure it is available to all users is increasingly difficult given the dynamic nature of representations.  

In this work, we survey three ongoing research threads that can inform web accessibility solutions: assessing web accessibility, modeling web user activity, and web application crawling. Current web accessibility research is continually focused on increasing the percentage of automatically testable standards, but still relies heavily upon manual testing for complex interactive applications. Along-side web accessibility research, there are mechanisms developed by researchers that replicate user interactions with web pages based on usage patterns. Crawling web applications is a broad research domain; exposing content in web applications is difficult because of incompatibilities in web crawlers and the technologies used to create the applications. We describe research on crawling the deep web by exercising user forms. We close with a thought exercise regarding the convergence of these three threads and the future of automated, web-based accessibility evaluation and assurance through a use case in web archiving. These research efforts provide insight into how users interact with websites, how to automate and simulate user interactions, how to record the results of user interactions, and how to analyze, evaluate, and map resulting website content to determine its relative accessibility.
\\
\\
\noindent \textbf{Approved for Public Release; Distribution Unlimited. Public Release Case Number 19-2337}



\end{abstract}






\maketitle

\section{Introduction}
\label{intro}

Accessibility of online content is becoming increasingly important as the government and many other organizations are moving critical functions online. Nearly 20\% of people in the U.S. have a disability \cite{census} and many of them may be affected by inaccessible websites. 

There are several barriers that are preventing web content owners from implementing good accessibility practices: lack of knowledge, lack of testing capability, and a lack of funding. There are many efforts attacking the first problem, either by creating accessibility standards (e.g., W3C \cite{wai}) or by creating training programs for testers and developers (i.e., Trusted Tester Program \cite{dhsttp}). While the Trusted Tester program is starting to work on creating experienced accessibility testers, its main focus is training non-technical workers to use automated tools. However, the automated tools they are taught to use are only capable of finding roughly 30\% of accessibility issues \cite{vigo2011}. As a result, a government agency or business may spend hundreds of thousands of dollars a year attempting to update its online systems to conform to accessibility standards and still have a system that is inaccessible to a portion of the population. 

Further, web applications that rely on JavaScript and Ajax to construct dynamic web pages pose further challenges for both disabled users and automated evaluation of accessibility. These applications are increasing in frequency on the web. Despite this increase, current automated tools are insufficient for mapping the states of dynamic web applications \cite{jbrunelleDissertation}, leading to undiscoverable accessibility violations.

As a result of these complexities, a universal method of automated assessments of web application accessibility is expensive and challenging -- if not impossible. Current methodologies involve human driven testing, or entirely manual analysis of accessibility.

Given the advancements and combination of web science and accessibility theory, it should be feasible to inform future ubiquitous, automated web accessibility tools.

\subsection{How the Web Works}
\label{webcrawling}
To begin, it is important to cover basic concepts surrounding web crawling and the web architecture. Universal Resource Identifiers (URIs) are generic forms of the more-familiar Universal Resource Locators (URLs). URIs identify \emph{resources} on the web; resources are abstract things with server-delivered \emph{representations}. For example, the resource for the Washington, D.C. weather report may have a HTML representation that allows web users to visually or graphically consume the report through the browser. Alternatively, the report may also have a textual or image representation. 

Users often ``visit'' the resource through the URI; this is a process called \emph{dereferencing} the URI and is often obfuscated from the user by the web browser. During the process of dereferencing a URI, the browser will issue a HTTP request for the resource to the server, and the server will respond with a HTTP response that includes the HTTP status code (e.g., 200 OK, 404 Not Found) along with a representation (when appropriate, such as delivering HTML with a 200 status code).

JavaScript is a client-side scripting language that (often) executes in a browser. Ajax is a technology that -- through JavaScript -- makes additional requests to dereference URIs to pull new representations into a web page. JavaScript -- and often Ajax -- are used to create web \emph{applications} as opposed to static representations. Web applications are meant to be interactive and often are referred to as Rich Internet Applications (RIAs). JavaScript is used to manipulate the Document Object Model (DOM)\footnote{The DOM is the tree structure of the HTML elements in a representation.} in an HTML representation. \emph{Deferred representations} rely on JavaScript -- and specifically Ajax --  to initiate requests for embedded resources after the initial page load \cite{jbrunelleDissertation}.

\subsection{Importance of Accessibility}
\label{accessibility}

As of 2010, about 15\% of the world's population (over 1 billion people) live with some form of disability and that number is growing as the population ages \cite{whodisability}.  Ensuring these individuals can participate and contribute to society requires accessible technology.  Section 508 \cite{section508} dictates accessibility standards for the US Government technology and electronic communications\footnote{As cited by Kanta, government accessibility varies widely \cite{govAccessibility}.}. Additional legal requirements also prevent discrimination of individuals with disabilities through technology \cite{dollaws, ada}. 

As a reflection of accessibility's importance to production web applications, Garrison recommends accessibility testing be incorporated into the software development process alongside security evaluations \cite{cicdAccessibility}. Ensuring web content and applications are accessible, prevents discrimination against disabled users. It ensures information is available for users.

There are many efforts focused on web accessibility such as accessibility standards (e.g., W3C \cite{wai}) and training programs for testers and developers (i.e., Trusted Tester Program \cite{dhsttp}). As a result, a government agency or business may spend hundreds of thousands of dollars a year attempting to conform to accessibility standards. Unfortunately, they may still fall short and have a system that is inaccessible to a portion of the population \cite{vigo2011}. 

\section{Related Work}
\label{relatedwork}

A wide variety of works can come together to inform the future of automated and higher quality web accessibility evaluations. In this section, we provide an overview of the relevant works in each domain. In subsequent sections, we describe how these should come together to create a viable framework for reducing the effort and increasing the quality of web accessibility analysis.

First, we discuss existing web accessibility work to include the existing guiding standards for website accessibility along with existing research that explores automatically identifying and quantifying accessibility (primarily informed by the existing standards). This analysis of prior works establishes the current state of the art in accessibility measurement and accessible web design.

Second, we present prior works in using web user behavior models -- both manually and automatically -- to interact with and discover data from web pages. This establishes the ways in which prior researchers have designed user interactivity models for the use of crawling and testing, and therefore informing a future model for disabled user models. 

Third, we summarize several works on automatically crawling web applications (e.g., those driven by Ajax) for both testing and evaluation purposes. These works can inform a model for mapping the states of a web application for accessibility evaluation; this may include the models for state equivalency.

Fourth, we describe ways in which the deep web (e.g., portions of web pages that are ``hidden'' behind forms or other user input mechanisms) can be automatically crawled. Future frameworks will need to crawl these portions of web pages, we present these materials as current methods that have informed our approach.

Fifth and finally, we describe a domain that already leverages each of the prior four research domains: using crawling for web archiving. The web archiving domain is reliant upon high recall crawling -- as is our accessibility analysis -- and additional models of state equivalency and techniques to improve coverage. As such, these adaptations of the testing, crawling, and user models domains are complemented by the web archiving crawler approaches, thus exemplifying the feasibility of applying these research areas to accessibility.

\subsection{Approaches for Web Accessibility Measures}
\label{accessibilitymeasure}

The need to measure or test for accessibility of websites is not new. In fact, standards  (primarily Section 508 and WCAG) exist for guiding and maximizing the accessibility of websites \cite{section508, wcag}. However, adherence to these standards is not uniform (despite the efforts of the W3.org Accessibility Conformance Task Force \cite{w3task}\footnote{The W3.org Accessibility Conformance Task Force has a goal to create a standardized framework for creating accessibility rules. These rules can be testing either automatically, semi-automatically, or manually.}), and identifying the points in a web application workflow at which users may have difficulty consuming web content is an outstanding challenge in the domain. Further, Hackett et al. showed that websites are becoming increasingly inaccessible over time \cite{Hackett2003}. 

Brajnik and Lomuscio created accessibility metrics for different categories of disabled users and compared their auto-generated metrics to human subject matter expert (SME) assessments \cite{braj2007}. Brajnick continued this work to determine which accessibility aspects are of the highest impact for disabled users \cite{braj2008}. Vigo and Brajnik provided an overview of the state of automated accessibility measures, identifying that high-impact accessibility violations are reliably detected but lower priority metrics are in varying degrees of agreement with SME assessments \cite{vigo2011}. Mankoff et al. described their tool for assessing accessibility of web pages for blind users and recommend an optimal method of blindness accessibility evaluation using a screen reader \cite{mankoff2005}. 

W3.org created techniques for adhering to WCAG guidelines \cite{w3wcag}. The techniques are development practices and are separated into three groups depending on how well they meet the guideline criteria. Note that these are informative, or suggestions to developers on best practices but are not required to meet guideline criteria.  This is an attempt to create a uniform set of techniques for adherence to WCAG standards. The techniques are left intentionally informative as not to stifle innovative techniques especially since they may change with new technology. A workshop run by Wild and Byrne-Haber at the ICT Accessibility Testing Symposium discussed several tools for accessibility testing and testing methodologies \cite{accessibilityTesting}. 

Kousabasis et al. identified requirements for assessing a web site's accessibility based on a website's design \cite{kouts2010}. 
Parmanto and Zeng provide a way to quantify the accessibility of a website using automated analysis of Section 508 and WCAG compliance. They also provide the accessibility of the site over time using the Internet Archives Wayback Machine\footnote{We discuss the intersection between accessibility, crawling, and archiving RIAs in Section \ref{archiving}.} \cite{iawebarchive, waybackarchives2}. Harper and Chen used the Wayback Machine to provide a longitudinal study of popular and random website accessibility using WCAG guidelines, finding that accessibility standards are met less than 10\% of the time in web pages, potentially due to the increased adoption of Ajax \cite{harper2012}.

Brown and Harper created the Ajax Time Machine to allow disabled users to roll-back version of web pages on demand when Ajax caused pages to change too frequently to be used \cite{ajaxtimemachine}. Carta et al. created a set of proxy-based tools to evaluate the usability of a web page using a pre-defined set of user models \cite{usabilityEval} and Oney and Myers presented a model for predicting user interactions on RIAs \cite{Oney:2009:FUI:1685992.1686124}. Borodin et al. presented a method of identifying aspects of web pages updated via Ajax to be presented to blind users \cite{Borodin:2008:WNM:1414471.1414499}.

Puppeteer is a Google product that offers a way to understand the accessibility violations of DOM elements \cite{puppeeer}. The General Services Administration (GSA) uses the pa11y tool \cite{pa11y} to crawl and assess the accessibility of static HTML and the HTML that results from scripted user interactions. The tool takes a list of user interactions to execute based on user instructions (e.g., ``click element A, click element B, ...''). However, it lacks the ability to automatically discover and interact with the DOM. JAWS a screen reader that is used in manual accessibility testing \cite{jaws}.

Ashok et al. created a model for enabling vision-impaired web users \cite{ashokAria, ashokWOZ}. The model allows content owners to create an annotated DOM based on JavaScript-generated user interfaces. The model then allows vision impaired users to interact with the widget with voice instructions. The approach showed improvement over screen-readers for users' ability to interact with JavaScript driven web applications. In an effort to generalize their approach, Ashok et al. explored ways to automatically classify user interface components based on their DOM elements \cite{ashokClassification}.

Several efforts have been performed to identify models of testing and have analyzed common violations or accessibility challenges with types of websites. Croniser identified several accessibility violations that occur when using a content management system such as WordPress \cite{wordpressAccessibility}. Wild described an analysis of common accessibility violations within social media pages(e.g., infinite scrolling and media autoplay) \cite{socialMediaAccessibility}.

The United States Access Board is focused -- among other things -- on developing and maintaining accessibility standards\footnote{\url{https://www.access-board.gov/research}}. As part of this, the board funds independent research on accessibility and accessible design. This board demonstrates the importance of accessibility with a focus on accessibility in general (e.g., to include vehicles, playgrounds), rather than a more narrow focus on web-based accessibility.

The efforts described in this section demonstrate the breadth of research efforts around accessibility analysis, and the approaches that prior researchers have built upon to achieve better accessibility results. The efforts include standards, regulations, models, and tools that have all evolved over time. The culmination being models that attempt to create semi-automated information consumption tools for disabled users.

\subsection{Web User Interaction Models}
\label{usermodels}

In order to automatically evaluate the accessibility of a web application, a user's interactions may need to be simulated. The methods by which users and their interactions are studied and simulated are very important to automating accessibility assessments. Of particular interest are events on the client that are driven by JavaScript or Ajax since these are often hidden from traditional web-based automation tools.

Researchers have developed tools to record interactions and the resulting DOM changes that occur or Ajax that is sent and requested from the client and server \cite{navigationReplay}. Bartoli et al. demonstrated the ability to record and replay navigations on Ajax-heavy resources \cite{ajaxReplay}. Their tool records interactions and replays the trace in an effort to test user interfaces with the goal of overcoming the difficulties in automated testing and programmatic interaction with client-side interfaces introduced by Ajax.

Similarly, FireCrystal \cite{Oney:2009:FUI:1685992.1686124} identified the interactive portions of a page and shows the associated code. It also records any events that occur within the code and interactive elements and identify any DOM changes that result from these events\footnote{This solution cannot work with Flash or other compiled languages on the client.}. VisualEvent \cite{seleniumpjs} uses a similar method to identify the various portions of a web page that have event listeners in JavaScript attached.

Rather than record interactions, Gorniak worked to predict future user actions by observing past interactions by other users \cite{Gorniak:2000:PFU:647288.721746}. This work modeled users as agents based on their interactions and clients based on potential state transitions. The agents would make the next most probable state transition, indicating what action the user is expected to take.

Leveraging user interaction patterns, Palmieri et al. generated agents to collect hidden web pages by identifying navigation elements from current selections \cite{autoAgents}. They represented the navigation patterns from these agents as a directed graph and would train the agents to interact with navigation elements. This includes forms, which would be filled out based on repository attributes. Drop-down menus would be randomly selected by the agents to spawn new pages. This method generated 80\% of all potential pages.

Fenstermacher and Ginsburg also noted that the majority of web mined data resides on servers and that data native to the client is not captured by crawlers \cite{Fenstermacher}. To begin recording user interactions with client-side representations, the authors developed an Internet Explorer framework for modeling and capturing client-side events. Their work is an early, yet limited, effort to capture data residing on the client\footnote{Interactions such as web form submissions are missed.}. 

Understanding and modeling user behavior can help uncover client-side states and events to more thoroughly test, debug, and interact with a page. Puerta and Eisenstein established a model for representing interaction data on the client \cite{Puerta:2002:XCR:502716.502763}. Webquilt \cite{Hong:2001:WFC:371920.372188, 2001:WPA:502115.502118, Waterson:2002:DTU:1556262.1556276} is a framework for capturing user interactions during a browsing session and subsequently visualizing the user's experience. Webquilt creates this functionality with a proxy logger, action inferencer, graph merger, graph layout, and visualization components. 

ActionShot is another add-on for browsers that automatically records user interactions for data mining, study, and replay \cite{heresWhatIDid}. This solution specializes in capturing client-side activity at fine-grained levels, such as click locations, targets, and resulting events.  Koala is a solution in which human-written natural language scripts can be played to automate processes in web applications \cite{koala}. This project is meant to facilitate often-used user interactions for user convenience. However, the scripts must be downloaded and played from a wiki.

Memex is a system that records user interactions with web pages \cite{memex}. It aims to develop a series of clickstreams for groups of similar interests. The system archives a search query from the user and also archives the sequence of pages a user visits. The system can allow for private and public browsing streams and public streams can be shared with users. 

This body of work demonstrates that a broad domain of work exists to study the ways in which users interact with web pages. The application of these works include:
\begin{itemize}
\item Predicting user interactions or activities 
\item Automated web application testing
\item Modeling information discovery behavior
\item Methods by which events are constructed on a web application
\end{itemize}

\subsection{JavaScript Web Application Crawling}
\label{jstesting}

Along with the importance of simulating and replicating user behavior, the ability to automatically recognize the interactive components of a web page is equally important. Assuming a tool or service is able to identify the interactive components of a web page, a user interactivity model may be implemented to simulate a fully-abled or disabled user interactivity model. 

Several efforts have studied client-side state with attention toward testing RIAs. Mesbah et al. performed several experiments regarding crawling and indexing representations of web pages that rely on JavaScript \cite{mesbahCrawling, mesbahInferState, mesbahDiss} focusing mainly on search engine indexing and automatic testing \cite{mesbahTesting, mesbah2}. Singer et al. developed a method for predicting how users interact with pages to navigate within and between web resources \cite{hyptrails}. Rosenthal described the difficulty of crawling representations reliant on JavaScript (and often for the purposes of web archiving) \cite{iipc2013, futureWeb}. Rosenthal et al. extended their LOCKSS\footnote{LOCKSS stands for ``Lots Of Copies Keep Stuff Safe'' and refers to the notion that many distributed copies of web content would ensure its preservation similar to redundant storage in data centers.} work to include a mechanism for handling Ajax \cite{dshrDlib}. Using CRAWLJAX and Selenium to automate events on DOM elements and tools to monitor the HTTP traffic, they captured Ajax-specific resources. Zheng and Zhang proposed an approach to perform bug tracking in Ajax applications \cite{ajaxBugs}, and Choudary et al. proposed a cross-browser approach for testing RIAs \cite{choudhary2}.

Mesbah et al. \cite{mesbah} also identified browser cookies and HTTP 404 responses as problems they encountered when trying to recreate states for a representation. As a result, their crawled pages are not sharable without special software installed on the user's machine. Duda's work with AJAXSearch \cite{duda2008} generated all possible states of a deferred representation and captured the generated content. Benjamin \cite{kamara2011} took the approach of providing a canonical method of crawling RIAs for search indexing purposes. 

Raj et al. performed crawls on Ajax-driven RIAs and detailed a method of using state transitions (defined by JavaScript events) and state equivalence (using the DOM trees for comparison) \cite{raj2018}. Li et al. took a different approach that uses a proxy to inject JavaScript into crawled pages \cite{li2018}. In their approach, they defined state transitions in the same way as Raj et al. Fejfar described an approach for crawling RIA DOM elements using a human-in-the-loop approach, allowing the human to direct the user interactions that should be crawled \cite{fejfar2018}.

A prime example from which both our current research as well as prior work is based is that by Dincturk et al. \cite{dincturkAjax, mappingState, dincturkDiss}. They present a model for crawling RIAs by constructing a graph of client-side states (they refer to these as ``AJAX states'' within the Hypercube model). Their work, which serves as the mathematical foundation and framework inspiration for our work, identifies the challenges with crawling Ajax-based representations and uses a \emph{hypercube strategy} to efficiently identify and navigate all client-side states of a deferred representation. Their model defines a client-side state as a state reachable from a URL through client-side events and is uniquely identified by the state's DOM\footnote{Other research exists on defining state outside of the URI. HTML5 also offers client-side storage to specify and store the client-side state \cite{w3cappstate}. Google provides client-side state storage that enables back-button functionality by keeping a stack of state changes \cite{rsh, uriDupes, uriDupes2}.}. That is, two states are identified as equivalent if their DOM (e.g., HTML) is directly equivalent.

From these efforts, one can observe a need for customized methods of automatically crawling JavaScript and Ajax-driven web applications. Traditional tools are unable to adequately execute JavaScript on a representation and -- as a result -- may miss important information on the page.

\subsection{Crawling the Deep Web}
\label{deepweb}

Past works have focused on crawling and discovering deep web resources (e.g., content requiring a user to log in to a site or fill out a form). Crawling pages with input forms is particularly difficult given the near-infinite combinations of input texts that can be constructed. User interaction models will likely require interactions with -- in part -- forms and other methods of free-text inputs.

Kundu and Rohatgi used an approach to generate potential input queries to web forms to uncover hidden web representations \cite{kundu2017}. Their approach generated the input queries through web page clustering and sampling using TF/IDF\footnote{TF/IDF stands for Term Frequency over Inverse Document Frequency and measures the uniqueness of a term or set of terms -- often in a single document -- in the context of a corpus.} and a random forest classifier to construct the input text. Raghavan and Garcia-Molina have proposed frameworks for uncovering the deep web behind input forms using a variety of input generation techniques \cite{ilprints456}. Lage et al. presented a set of user models that interact with web forms to uncover hidden content \cite{autoAgents}. 

He et al. worked to sample and index  the deep web \cite{accessingDeepWeb}. Yeye utilized server query logs and URL templates to discover deep content \cite{deepWebCrawl}. This prototype successfully collected the RIAs and associated states. Likarish and Jung crawled the deep web to create a collection of malicious JavaScript using Heritrix and comparing captured scripts to blacklisted content \cite{Likarish:2009:TWC:1651309.1651317}. Ntoulas et al. captured the textual hidden web with a guess-and-check method of supplying server- and client-side parameters \cite{dl_Web_content}.

In contrast, Google's search engine crawlers -- as of 2014 executed JavaScript to discover new URIs to crawl but does not perform page interactions \cite{googleJS}. Content owners are able to use HashBang URIs \cite{hashbang, hashjen, googleAjax} to give Google crawlers the ability to locate and crawl the page, and also retrieve the client-side state \cite{googleJS, googleCrawlAjax, journals/corr/abs-0909-1785, hiddenurls}. This effectively provides a JavaScript -- and potentially Ajax -- driven representation in static form for a crawler to discover and index. Similarly, Google proposed (and subsequently created) a headless browsing solution to query, interact, and render the content of the page for crawling purposes \cite{googleCrawlAjax2}.

There are various methods to simulate user input into a web form. However, they vary in their ability to be automated and the level to which entered information/data must be constructed.

\section{A Use Case: Web Application Crawling for Archiving}
\label{archiving}

The various research threads we have presented in this paper so far are applied within a single domain despite their unique challenges and research directions. Web archiving has a similar combination of challenges to web accessibility testing, particularly when attempting to automate both activities. 

\subsection{What is Web Archiving?}
Web archiving is a domain that is increasing in popularity and importance, particularly given the increased attention in today's society being paid to information availability. Recent articles in \emph{The New Yorker} \cite{newyorker} and \emph{The Atlantic} \cite{lafrance} provide an overview of how web archives are playing an important role in information provenance and accountability. 

A primary driving reason that web archives are relevant, important, and face a challenging mission is that \emph{web resources are ephemeral}, existing in the perpetual ``now''. Important historical events frequently disappear from the web without being preserved or recorded. We may miss pages because we are not aware they should be saved or because the pages themselves are hard to archive. 

\subsection{Web Archiving Motivation} 
On July 17, 2015, Ukrainian separatists posted a video and associated claim on VK (VKonkakte, a Russian social media site) that they shot down a military plane in Ukrainian airspace. 
 The following investigation revealed that the plane was more likely the commercial Malaysian Airlines Flight 17 (MH17). The Ukrainian separatists removed their claim after the investigation revealed that their target was a commercial airliner rather than a military aircraft. The Internet Archive \cite{iawebarchive}, using Heritrix \cite{heritrix, Sigurosson:Incremental-Heritrix}, was able to archive the claimed credit for downing the aircraft; this is evidence that Ukrainian separatists shot down MH17 \cite{csm}. This (among others \cite{ainsworthPop}) is an example of the importance of high-fidelity web archiving to record history and establish evidence of information published on the web.

However, not all historical events are archived as fortuitously. In an attempt to limit online piracy and theft of intellectual property, the U.S. Government proposed the Stop Online Piracy Act (SOPA) \cite{sopaabc}. While proposal and ultimate failing of SOPA may be a minor footnote in history, the protest in response is significant. On January 18, 2012, many websites organized a world-wide blackout in protest of SOPA. Wikipedia was one such site that blacked out their site by using JavaScript to load a ``splash page'' that prevented access to the site's content\footnote{Extended examples and figures are omitted in this paper but can be found in prior works \cite{sopapost, brunelleSopa}.}.

The Internet Archive, again using Heritrix, archived the Wikipedia site during the protest. However, the archived January 18, 2012 page, as replayed in the Wayback Machine \cite{waybackarchives2}, does not include the splash page. This is because archival crawlers such as Heritrix do not execute JavaScript during crawling and were unable to archive the splash page in real-time. Wikipedia's protest as it appeared on January 18, 2012 has been lost from the archives and  -- had it not been for human efforts -- would be lost from history. 

Unlike the MH17 example (which establishes our need to archive with high fidelity), the SOPA example is not well represented in the archives\footnote{Even though it does not execute JavaScript, Heritrix v. 3.1.4 does peek into the embedded JavaScript code to extract links where possible \cite{htrixJS}. In contrast, Google's search engine crawlers execute JavaScript to discover new URIs to crawl but does not perform page interactions \cite{googleJS}.}. This is a result of browsers implementing (and content authors adopting) client-side technologies such as JavaScript faster than web crawler developers can adapt to handle the new technologies. This leads to a difference between the web that crawlers can discover and the web that human users experience -- a challenge impacting the archives as well as automated mechanisms of uncovering content (such as the content that should be evaluated for accessibility). 

\subsection{Prior works in Web Application Archiving}

Over time, web resources have increasingly used JavaScript to load embedded resources \cite{ijdl}. Automatically testing, crawling, and evaluating representations of web resources that leverage JavaScript to load embedded resources and build representations is challenging, and often requires special tools beyond traditional crawlers. As such, the domain of web archiving is rich with examples of RIA crawling approaches. For example, web archiving crawlers must leverage specialized tools to accurately archive representation that use JavaScript (e.g., Heritrix \cite{Sigurosson:Incremental-Heritrix} using specialized crawling techniques to ``peek'' into JavaScript to understand what it will do \cite{djangoPJS, htrixJS}) but often fail to perform adequately \cite{cnn}. 

Browsertrix \cite{browsertrix} and WebRecorder.io \cite{webrecorder} are page-at-a-time archival tools for deferred representations and descendants, but they require human interaction and are not suitable for automated web archiving. Archive.is \cite{archivetoday} handles deferred representations well, but is a page-at-a-time archival tool and must strip out embedded JavaScript to perform properly, making the resulting representation non-interactive. Umbra is a tool used to crawl a human curated set of pre-defined URI-Rs \cite{umbra} and Brozzler uses headful or headless crawling for archiving \cite{brozzler}.

Our prior work showed that specialized crawling can expose the JavaScript-dependent events and deferred representations \cite{brunellejcdl2018} using Heritrix and PhantomJS \cite{pjs, crawlingDeferred}. This work was adapted from the research by Dincturk et al. \cite{dincturkAjax} to construct a model for crawling RIAs by discovering all possible descendants and identifying the simplest possible state machine to represent the states. In our prior work, we define state transitions as user interactions (and that trigger JavaScript event listeners) and state equivalence is determined by the set of embedded resources required to build out the page. This differs from other researchers' use of strict DOM or DOM tree equivalency to define a state. 

The web archiving domain and interpretations of the Dinctruk et al. Hypercube model from the application testing domain inspire approaches automatically crawling web pages in the web accessibility domain.

\subsection{The Influence of Web Archiving}

Based on the models being researched and implemented in the web archiving domain and the need for automated accessibility tools and approaches, a confluence of approaches and technologies must be employed. These research methods used to discover, crawl, map, and execute interactions on web applications can be adapted for accessibility. For each representation, the interactive components must be discovered, mapped, exercised, and evaluated with user interactivity models. The research being performed and integrated at the center of web archiving is a potential roadmap for performing the same work within automated web-based accessibility evaluations.

\section{Our Model for Accessibility Testing}
\label{ourwork}

In our current work, we are using these efforts as inspiration for a model that automatically detects interactive elements in a web-based representation and uses user models to navigate a web application. We are creating user models that have various disabilities and -- therefore -- limit the simulated users' ability to exercise UI elements in various ways. For example, a user may be unable to use the mouse and is restricted to keyboard interactions. 

Our intent is to create a method of automatically crawling and assessing the accessibility of a web application based on the client-side events the user model is able to trigger. We will compare the navigation of a fully-abled user model with that of a disabled user model to compare what states of a web application are navigable for a fully-abled user versus a user with a disability.

We anticipate this work being used to enable faster, less expensive, and more accurate accessibility testing for US government organizations and ultimately informing future accessibility testing efforts. We are relying on the prior research in accessibility, user models, web crawling, and are using web archiving research as our template for implementing our work.

\section{Conclusions}

In this paper, we have provided a selected literature review of research threads (i.e., assessing web accessibility, modeling web user interactions, and crawling web applications) that can inform automated web-based accessibility evaluations. There is a need to reduce the manual and human-driven effort being dedicated to evaluating the accessibility of websites, if they are evaluated at all. Creating a service or tool that helps with these assessments given the challenges unique to web applications would be a major advancement toward assessing web accessibility. 

We recommend further research be performed on the intersection of the research threads presented in this paper and demonstrate an applicable convergence of these research methods by discussing their impact on web archiving. Our goal is to stimulate the development of technology that improves accessibility of web information to disabled users.

\section{Acknowledgments}
We would like to thank Sanith Wijesinghe, the Innovation Area Lead funding this research effort as part of MITRE's internal research and development program (the MIP). We also thank the numerous collaborators that have assisted with the maturation of our research project.

\copyright 2019 The MITRE Corporation. All Rights Reserved.

Approved for Public Release; Distribution Unlimited. Case Number 19-2337.


\bibliographystyle{abbrv}
\bibliography{_mybibtex}

\end{document}